\documentclass[letterpaper]{www2004-submission}

\usepackage{times}
\begin{document}

\title{Practical Semantic Analysis of Web Sites and Documents}

\numberofauthors{1}

\author{
\alignauthor Thierry Despeyroux\\
       \affaddr{I.N.R.I.A. - Rocquencourt}\\
       \affaddr{AxIS Group}\\
       \affaddr{B.P. 105 - 78153 Le Chesnay Cedex, France}\\
       \email{thierry.despeyroux@inria.fr}
}

\maketitle

\begin{abstract}

As Web sites are now ordinary products, it is necessary to explicit
the notion of quality of a Web site.  The quality of a site may be
linked to the easiness of accessibility and also to other criteria
such as the fact that the site is up to date and coherent. This last
quality is difficult to insure because sites may be updated very
frequently, may have many authors, may be partially generated and in
this context proof-reading is very difficult.  The same piece of
information may be found in different occurrences, but also in data or
meta-data, leading to the need for consistency checking.

In this paper we make a parallel between programs and Web sites.  We
present some examples of semantic constraints that one would like to
specify (constraints between the meaning of categories and
sub-categories in a thematic directory, consistency between the
organization chart and the rest of the site in an academic site).  We
present quickly the Natural Semantics~\cite{Kahn87,Des87}, a way to
specify the semantics of programming languages that inspires our
works. 
Then we propose a specification language for semantic
constraints in Web sites that, in conjunction with the well known
``make'' program, permits to generate some site verification tools by
compiling the specification into Prolog code. We apply our method to a
large XML document which is the scientific part of our institute
activity report, tracking errors or inconsistencies and also
constructing some indicators that can be used by the management of the
institute.

\end{abstract}

{Logic Programming}
{Document Preparation}[Languages and systems,Markup Languages]



\section{Introduction}

Web sites can be seen as a new kind of everyday product which is
subject to a complex cycle of life: many authors, frequent updates or
redesigns. As for other ``industrial'' products, one can think of the
notion of quality of a Web site, and then look for methods to achieve
this quality.

Until now, the main effort developed in the domain of the Web is
called the Semantic Web~\cite{Lee98,Lee2001}.  Its goal is to ease a
computer based data mining, formalizing data which is most of the time
textual. This leads to two directions:

\begin{itemize}

\item Giving a syntactical structure to documents.  This is achieved
with XML, DTDs, XML-schema, style sheets and XSLT~\cite{W3C}.  The
goal is to separate the content of Web pages from their visual
appearance, defining an abstract syntax to constrain the information
structure. In this area, there is, of course, the work done by the W3C,
but we can also mention tools that manipulate XML documents, taking
into account the conformance to DTD: Xduce~\cite{HP00} and
XM-$\lambda$~\cite{MS99}. This means that by using these languages we
know that the documents which are produced will conform to the specific
DTDs that they use, which is not the case when one uses XSLT.

\item Annotating documents to help computerized Web mining.  One will
use ontologies with the help of RDF~\cite{W3C,N3}, RDF-Schema or
DAML+OIL~\cite{DAMLOIL}. The goal is to get a computer dedicated
presentation of knowledge~\cite{Harmelen99b,Ontobroker,On2broker} to
check the content of a page with the use of
ontologies~\cite{Harmelen99a} or to improve information retrieval.

\end{itemize}

Our approach is different as we are concerned in the way Web sites are
constructed, taking into account their development and their
semantics. In this respect we are closer to what is called content
management.  To do that, we use some techniques coming from the world
of the semantics of programming languages and from software
engineering.

Web sites, as many other types of information systems, contain
naturally redundant information. This is in part due to accessibility
reasons.  Web pages can be annotated and the same piece of information
can exist in many different forms, as data or meta-data. We have to
make sure that these different representations of the same piece of
knowledge are consistent. Some parts of a Web site can also be
generated, from one or more databases, and again, one have to make
sure of the consistency between these databases and the rest of the
site.

Obviously, traditional proof-reading (as it may be done for a book)
is not possible for Web sites. A book may contain a structure
(chapters, sections, etc.) and references across the text, but its
reading can be essentially linear. An information system such as a Web
site looks more like a net.

We propose to apply techniques from software engineering to
increase the quality level of Web sites. In the context of
Web sites, it is not possible to make proofs as it is the case for
some sorts of programs or algorithms, but we will see that some
techniques used in  the area  of formal semantics of programming
languages~\cite{Gunter} can be successfully used in this context.

The work presented in this paper is limited to static Web sites and
documents. It can be easily extended to more general information
systems as many of them provide a Web interface or at least can
generate XML documents, answering to the two questions that we try to
solve: How can we define verification tools for Web sites an more
generally information systems? How can we mechanize the use of these
tools?

In a first section we will make a parallel between programs and Web
sites. In the second one we will show some examples of semantics
constraints in Web sites. Then we will explore the Natural Semantics
approach and will see how to adapt this method to our current problem.
We will finish this paper by describing experiments that have been
done and some implementation notes.

\section{From programs to web sites}

To execute a program you have, most of the time, to use a compiler
which translates the source code to executable code, unless you are
the end-user and someone else did this for you.

The goal of a compiler is not only to generate object code but also to
verify that the program is legal, i.e., that it follows the {\it
static semantics} of the programming language. For example, in many
programming languages one can find some declarative parts in which
objects, types and procedures are defined to be used in some other
places in statements or expressions. One will have to check that the
use of these objects is compatible with the declarations.

The static semantics is defined by opposition to the {\it dynamic
semantics} which express the way a program is executed. The static
semantics express some constraints that must be verified before a
program is executed or compiled.

A particularity of such constraints is that they are not local but
{\it global}: they may bind distant occurrences of an object in a
unique file or in many different files. A second particularity is that
these constraints are {\it not context-independent}: an
``environment'' that allows us to know what are the visible objects at
a particular point in the program is necessary to perform the
verifications.

Global constraints are defined by opposition to local constraints. As
a program may be represented by a labeled tree, a local constraint is
a relation between a node in this tree and its sons.  For example, if
we represent an assignment, the fact that the right hand part of an
assignment must be an expression is a local constraint.  On the other
hand, the fact that the two sides of an assignment must have
compatible types is a global constraint, as we need to compute the
types of both sides using an environment.

Local constraints express what is called the (abstract) syntax of the
programming language and global constraints express what is called its
static semantics. The abstract syntax refers to the tree
representation of a program, its textual form is called its concrete
syntax.

Representing programs by means of a tree is quite natural. Using a
B.N.F. to describe the syntax of a programming language gives already
a parse tree. Most of the time this tree can be simplified to remove
meaningless reduction level due to precedence of operators.  The
grammar rules express local type constraints on the parse tree and
explain what is a syntacticly correct program.  A straight way of
representing a program in a language like Prolog is to use (completely
instantiated, i.e., with no logical variables) terms, even if Prolog
terms are not typed.

The following example shows how a statement can be represented as a
Prolog term.

\begin{verbatim}
A := B + (3.5 * C);
\end{verbatim}

\begin{verbatim}
assign(var('A'), 
       plus(var('B'),
            mult(num(3.5),var('C'))))
\end{verbatim}

A Prolog term can itself be represented into XML as shown in the
following example:

\begin{verbatim}
<assign>
  <var name=''A''/>
  <plus>
    <var name=''B''/>
    <mult>
      <num value=``3.5''/>
      <var name=''C''/>
    </mult>
  </plus>
</assign>
\end{verbatim}

By defining a DTD for our language, we can constrain XML to allow only
programs that respect the syntax of the programming language.

\begin{verbatim}
<!ELEMENT assign (var, (plus|mult|num...))>
\end{verbatim}

This is equivalent to giving a signature to the Prolog terms

\begin{verbatim}
assign : Var x Exp -> Statement
\end{verbatim}

where Var and Exp are types containing the appropriate elements.

However, using types or DTDs it is not possible to express semantic
constraints, as for example ``The types of the left and right hand
sides must be compatible''. If the variable {\tt A} has been declared
as an integer, the statement given as example is not legal.

Web sites (we define a Web sites simply as a set of XML pages) are
very similar to programs.  In particular, they can be represented as
trees, and they may have local constraints expressed by the means of
DTDs or XML-shemas. As HTML pages can be translated to XHTML, which is
XML with a particular DTD, it is not a restriction to focus only to
XML web sites.

There are also differences between Web sites and programs:

\begin{itemize}

\item Web sites can be spread along a great number of files. This is the 
case also for programs, but these files are all located on the
same file system. With Web sites we will have to take into account that we 
may need to access different servers. 


\item The information is scattered, with a very frequent use of 
forward references. A forward reference is the fact that an object (or
a piece of information) is used before it as been defined or declared.
In programs, forward references exist but are most of the time limited
to single files so the compiler can compile one file at a time. This
is not the case for Web sites and as it is not possible to load a
complete site at the same time, we have to use other techniques.

\item The syntax can be poorly, not completely or not at all formalized,
with some parts using natural languages.

\item There is the possibility to use ``multimedia''. 
In the word of programs, there is only one type of information to
manipulate: text (with structure), so let's say terms. If we want to
handle a complete site or document we may want to manipulate images
for example to compare the content of an image with a caption.

\item The formal semantics is not imposed by a particular programming
language but must be defined by the author or shared between authors
as it is already the case for DTDs. This means that to allow the 
verification of a Web site along its life one will have to define
what should be checked. 

\item We may need to use external resources to define the static
semantics (for example one may need to use a thesaurus, ontologies or
an image analysis program). In one of our example, we call the {\tt
wget} program to check the validity of URLs in an activity report.

\end{itemize}

Meta-data are now most of the time expressed using an XML syntax, even
if some other concrete syntax can be used (for example the N3
notation~\cite{N3} can be used as an alternative to the XML syntax of
RDF). So, from a syntactic point of view, there is no difference
between data and meta-data.  Databases also use XML, as a standard
interface both for queries and results.

As we can see, the emergence of XML gives us a uniform framework for
managing heterogeneous data. Using methods from software engineering
will allow a webmaster or an author to check the integrity of its
information system and to produce error messages or warnings when
necessary, provided that they have made the effort to formalize the
semantics rules.

However, this is not yet completely sufficient.  Every programmer can
relate to some anecdote in which somebody forgets to perform an action
as recompiling a part of a program, leading to an incoherent
system. Formalizing and mechanizing the management of any number of
files is the only way to avoid such misadventure.

Again, at least one solution already exists. It is a utility program
named ``make''~\cite{stallman00gnu}, well known to at least unix
programmers. This program can manage the dependencies between files
and libraries, and it can minimize the actions (such as the calls to
the compilers) which must be done when some files have been modified.
However, this program can only managed files located on the same file
system and must be adapted to handle URLs when a Web site is scattered
over multiple physical locations.

\section{Examples of semantic\\constraints}

For a better understanding of the notion of semantic constraints we
will now provide two examples. More examples can be found in the
applications section.

\begin{itemize}

\item A thematic directory is a site in which documents or
external sites are classified. An editorial team is in charge of
defining the classification. This classification shows as a tree of
topics and subtopics.  The intentional semantics of this
classification is that a subtopic ``makes sense'' in the context of
the upper topic, and this semantics must be maintained when the site
is modified.

To illustrate this, here is an example:\\

{\bf Category}: Recreation\\
{\it Sub-Category}: Sports, Travel, Games, Surgery, Music, Cinema\\

The formal semantics of the thematic directory uses the semantics of
words in a natural language. To verify the formal semantics, we need
to have access to external resources, maybe a thesaurus, a
pre-existing ontology or an ontology which is progressively constructed
at the same time as the directory; how to access to this external
resource is not of real importance, the important point is that it
can be mechanized.

In the former example, the sub-category ``Surgery'' is obviously a
mistake.\\

\item An academic site presents an organization: its structure, its
management, its organization chart, etc. Most of the time this
information is redundant, maybe even inconsistent. As in a program,
one have to identify which part of the site must be trusted: the
organization chart or a diary (which is supposed to be up to date) and
can be used to verify the information located in other places, in
particular when there are modifications in the organization. The
``part that can be trusted'' can be a formal document such as an
ontology, a database or a plain XML document.

\end{itemize}

The issue of consistency between data and meta-data, or between many
redundant data appears in many places, as in the following
examples.

\begin{itemize}

\item Checking the consistency between a
caption and an image; in this case we may want to use linguistic tools
to compare the caption with annotations on the image, or to use image
recognition tools. 

\item Comparing different versions of
the same site that may exist for accessibility reasons. 

\item Verifying the consistency between a request to a database, with the
result of the request to detect problems in the query and verify the
plausibility of an answer (even when a page is generated from a
database, it can be useful to perform static verifications, unless the
raw data and the generation process can be proved, which is most of the
time not the case).  

\item Verifying that an annotation (maybe in RDF) is still valid when an
annotated page is modified.

\end{itemize}

\section{Formalizing Web sites}

As seen earlier, Web sites (and other information systems) can be
represented as trees, and more precisely by typed terms, exactly as it
is the case for programs. It is then natural to apply the formal
methods used to define programming languages to Web sites.

In our experiments, we have used the Natural
Semantics~\cite{Des87,Kahn87} which is an operational semantics derived
from the operational semantics of Plotkin~\cite{SOS} and inspired by
the sequent calculus of Gentzen~\cite{Gen69}.  One of the advantages
of this semantics is that it is an executable semantics, which means
that semantic definitions can be compiled (into Prolog) to generate
type-checkers or compilers. Another advantage is that it allows the
construction of proofs.

In Natural Semantics, the semantics of programming languages are
defined using inference rules and axioms. These rules explain how to
demonstrate some properties of ``the current point'' in the program,
named subject, using its subcomponents (i.e., subtrees in the tree
representation) and an environment (a set of hypothesis).

To illustrate this, the following rule explains how the statements
part of a program depends on the declarative one:
$$
\emptyset \vdash Decls \rightarrow \rho \qquad  \rho \vdash Stmts
\over
\vdash \textbf{declare } Decls \textbf{ in } Stmts
$$

The declarations are analyzed in an empty environment $\emptyset$,
constructing $\rho$ which is a mapping from variable names to declared
types. This environment is then used to check the statements part, and
we can see that in the selected example the statements part do not
alter this environment.

These inference rules can be read in two different ways: if the upper
part of the rule has been proved, then we can deduce that the lower
part holds; but also in a more operational mode, if we want to prove
the lower part we have to prove that the upper part holds.

The following rule is an axiom. It explains the fact that in order to
attribute a type to a variable, one has to access the environment.

$$
\rho \vdash  \textbf{var }X:T \qquad  \{X:T\} \in \rho 
$$

``Executing'' a semantic definition means that we want to build a
proof tree, piling up semantic rules which have been instantiated with
the initial data.

First, we have made experiments using directly Natural
Semantics~\cite{DT00,DT01}. These experiments showed that this style of
formal semantics perfectly fits our needs, but it is very heavy for
end-users (authors or webmaster) in the context of Web sites. Indeed,
as we can see in the former rules, the recursion is explicit, so the
specification needs at least one rule for each syntactical operator
(i.e., for each XML tag). Furthermore, managing the environment can be
tedious, and this is mainly due to the forward declarations, frequent
in Web sites (Forward declarations means that objects can be used
before they have been defined. This implies that the verifications
must be delayed using an appropriate mechanism as coroutine, or
using a two-pass process).

It is not reasonable to ask authors or webmasters to write Natural
Semantics rules. Even if it seems appropriate for semantic checking,
the rules may seem too obscure to most of them. Our strategy now is to
specify a simple and specialized specification language that will be
compiled in Natural Semantics rules, or more exactly to some Prolog
code very close to what is compiled from Natural Semantics.

We can make a list of requirements for this specification language:

\begin{itemize}

\item No explicit recursion.

\item Minimizing the specification to the points of interest only.

\item Simple management of the environment.

\item Allowing rules composition (points of view management).

\item Automatic management of forward declarations.

\end{itemize}

In a second step we have written various prototypes directly in
Prolog. The choice of Prolog comes from the fact that it is the
language used for the implementation of Natural Semantics. But it is
indeed very well adapted to our needs: terms are the basic objects,
there is pattern matching and unification.

After these experiments, we are now designing specification
language. Here are some of the main features of this language:

\begin{itemize}

\item Patterns describe occurrences in the XML tree. 
We have extended the language XML with logical variables. In the
following examples variable names begin with the {\tt \$} sign.

\item A local environment contains a mapping from names to values. 
The execution of some rules may depend on these values. A variable can
be read ({\tt =}) or assigned ({\tt :=}).

\item A global environment contains predicates
which are assertions deduced when rules are executed.  The
syntax of these predicates is the Prolog syntax, but logical variables
are marked with the {\tt \$} sign. The sign {\tt =>} means that its
right hand side must be added to the global environment.

\item Tests enable us to generate warnings or error messages when some
conditions do not hold. These tests are just now simple Prolog
predicates.  The expression {\tt ? pred / error} means that if the
predicate {\tt pred} is false, the error message must be issued.
The expression {\tt ? pred -> error} means that if the
predicate {\tt pred} is true, the error message must be issued.

\end{itemize}

Semantic rules contain two parts: the first part explains when the
rule can be applied, using patterns and tests on the local
environment; the second part describes actions which must be executed
when the rule applies: modifying the local environment (assignment),
adding of a predicate to the global environment, generating a test.

Recursion is implicit. It is also the case for the propagation of the two
environments and of error messages.

In the following section, we present this language with more details.

\section{A specification language to\\define the semantics of Web\\sites}

We give in this section a complete description of our specification
language.  The formal syntax is described in Appendix A.

\subsection{Patterns}

As the domain of our specification language is XML document, XML
elements are the basic data of the language. To allow
pattern-matching, logical variables have been added to the XML
language (which is quite different from what is done in XSLT).

Variables have a name prefixed by the {\tt \$} sign, for example:
{\tt \$X}. Their also exists anonymous variables which can be used when
a variable appears only once in a rule: {\tt \$\_}. For syntactical
reasons, if the variable appears in place of an element, it should be
encapsulated between {\tt <} and {\tt >}~: {\tt <\$X>}.

In a list, there is the traditional problem to know if a variable
matches an element or a sublist. If a list of variables matches a list
of elements only the last variable matches a sublist. So in 
\begin{verbatim}
<tag> <$A> <$B> </tag>
\end{verbatim}
the variable {\tt \$A} matches the first element
contained in the body of {\tt <tag>} and {\tt \$B} matched the rest of
the list (which may be an empty list).

When a pattern contains attributes, the order of the attributes is not
significant. The pattern matches elements that contain at least the
attributes present in the pattern.

For example, the pattern 
\begin{verbatim}
<citation year=$Y><$T><$R> </citation>
\end{verbatim}

matches the element
\begin{verbatim}
<citation type=''thesis'' year=``2003''>
  <title> ... </title> 
  <author> ... </author>
  ... 
  <year> ... <year>
</citation>
\end{verbatim}

binding {\tt \$Y} with ``2003'' 

and {\tt \$T} with {\tt <title> ... </title>}. 
The variable {\tt \$R} is bound to the rest of the list
of elements contained in {\tt <citation>}, i.e., the list of elements
\begin{verbatim}
  <author> ... </author>
  ... 
  <year> ... <year>
\end{verbatim}

When this structure is not sufficient to express some configuration
(typically when the pattern is too big or the order of some elements
is not fixed), one can use the following ``contains'' predicate:

\begin{verbatim}
<citation> <$A> </citation>
&  $A contains  <title> <$T> </title>
\end{verbatim}

In this case the element {\tt <title>} is searched everywhere in the
subtree {\tt \$A}.

\subsection{Rules}

A rule is a couple containing at least a pattern and an action.
The execution of a rule may be constrained by some conditions. These
conditions are tests on values which have been computed before and are
inherited from the context. This is again different from XSLT in which
it is possible to get access to values appearing between the top of
the tree and the current point. Here these values must explicitly be stored
and can be the result of a calculus.

In the following rule, the variable {\tt \$A} is bound to the year of
publishing found in the {\tt <citation>} element while {\tt \$T} is
found in the context.

\begin{verbatim}
<citation year=$Y > <$A> </citation>
  & currentyear = $T
\end{verbatim}

The effect of a rule can be to modify the context (the modification is
local to the concerned subtree), to assert some predicates (this is
global to all the site) and to emit error messages depending on tests.

The following example sets the value of the current year to what is found
in the argument of the {\tt <activityreport>} element.

\begin{verbatim}
<activityreport year=$Y > 
   <$A> 
</activityreport>
  =>  currentyear := $Y ;
\end{verbatim}

The two following rules illustrate the checking of an untrusted part
of a document against a trusted part. 

In the trusted part, the members of a team are listed (or declared).
The context contains the name of the current team, and for each person
we can assert that it is a member of the team.

\begin{verbatim}
<pers firstname=$F lastname=$L> 
   <$_> 
</pers>
  & teamname = $P  
  =>  teammember($F,$L,$P) ;
\end{verbatim}

In the untrusted part, we want to check that the authors of some
documents are declared as members of the current team.  If it is not
the case, an error message is produced.

\begin{verbatim}
<author firstname=$F lastname=$L> 
   <$_> 
</author>
  & teamname = $P  
  ?  teammember($F,$L,$P) /
     <li> Warning: <$F> <$L> is not 
          member of the team 
          <i> <$P> </i>
     </li>;
\end{verbatim}

\subsection{Dynamic semantics}

This section explains how semantic rules are executed.  

On each file, the XML tree is visited recursively. A local
environment which is initially empty is constructed. On each node, a
list of rules which can be applied (the pattern matches the current
element and conditions are evaluated to true) is constructed, then
applied. This means that all rules which can be applied are evaluated
in the same environment. The order in which the rules are then applied
is not defined.

During this process, two results are constructed. The fist one is a
list of global assertions, the second is a list of tests and related
actions. These are saved in files.

A global environment is constructed by collecting all assertions
coming from each different environment files. Then using this global
environment, tests are performed, producing error messages if there
are some.

To produce complete error messages, two predefined variables exist:
{\tt \$SourceFile} and {\tt \$SourceLine}. They contain respectively,
the name of the current file which is analyzed and the line number
corresponding to the current point. This is possibly due to the use of
a home made XML parser which is roughly described in Appendix A.

\section{Applications}

\subsection{Verifying a Web site}

The following example illustrates our definition language. We want to
maintain an academic site. The current page must be trusted and
contains a presentation of the structure of an organization.

\begin{verbatim}
<department> 
 <deptname><$X></deptname> 
 <$_>
</department>   
   =>   dept:=$X
\end{verbatim}

The left hand side of the rule is a pattern. Each time the pattern is
found, the local environment is modified: the value matched by the
logical variable {\tt \$X} is assigned to the variable {\tt
dept}. This value can be retrieved in all the subtrees of the current
tree, as in the following example. {\tt \$\_} matches the rest of what
is in the {\tt department} tag.

Notice that, unlike what happens in XSLT in which it is possible to
have direct access to data appearing between the root and the current
point in a tree, we have to store explicitly useful values in the
local environment. In return one can store (and retrieve) values which
are computed and are not part of the original data, and this is not
possible with XSLT.

\begin{verbatim}
<head><$P></head>
  &  dept=$X   
  =>   head($P,$X)
\end{verbatim}

We are in the context of the department named {\tt \$X}, and {\tt \$P}
is the head of this department. {\tt head(\$P,\$X)} is an assertion
which is true for the whole site, and thus we can add this assertion
to the global environment. This is quite equivalent to building a
local ontology.  If, in some context, an ontology already exists, we
can think of using it as an external resource. Notice that a triplet
in RDF can be viewed as a Prolog predicate~\cite{Peer2002}.

\begin{verbatim}
<agent><$P></agent> 
  ? appointment($P,$X) 
    / <li> Warning: <$P> line <$SourceLine>
           does not appear in the current
           staff chart.
     </li>;
\end{verbatim}

This rule illustrates the generation of error messages. In the XML
text, the tag {\tt agent} is used to indicate that we have to check
that the designated person has got an appointment in a department. The
treatment of errors hides some particular techniques as the generated
code must be rich enough to enable to locate the error in the source
code.

\subsection{Verifying a document and inferring new data}

As a real sized test application, we have used the scientific part of
the activity reports published by Inria for the years 2001 and 2002
which can be found at the following URLs:\\
http://www.inria.fr/rapportsactivite/RA2001/index.html and\\
http://www.inria.fr/rapportsactivite/RA2002/index.html.

The sources of these activity reports are LaTex documents, and are
automatically translated into XML to be published on the Web.

The XML versions of these documents contain respectively 108 files and
125 files, a total of 215 000 and 240 000 lines, more than 12.9 and
15.2 Mbytes of data. Each file is the reflect of the activity of a
research group.  Even if a large part of the document is written in
French, the structure and some parts of the document are
formalized. This includes parts speaking of the people and the
bibliography.

The source file of our definition can be found in Appendix B.

Concerning the people, we can check that names which appears in the
body of the document are ``declared'' in the group members list at the
beginning. If it is not the case, the following error message is produced:

\begin{verbatim}
Warning: X does not appear in the list of 
project's members (line N)
\end{verbatim}

Concerning the bibliography of the group, the part called
``publications of the year'' may produce error messages like the
following one:

\begin{verbatim}
Warning: The citation line 2176 has not been
published during this year (2000)
\end{verbatim}

We have also use Wget to check the validity of URLs used as citation,
producing the following error messages:

\begin{verbatim}
Testing of URL 
http://www.nada.kth.se/ruheconference/
line 1812 in file ``aladin.xml" replies: 
http://www.nada.kth.se/ruheconference/: 
14:50:33 ERREUR 404: Not Found.

Testing of URL 
http://citeseer.nj.nec.com/ning93novel.html
line 1420 in file "a3.xml" replies: 
No answer or time out during wget,  
The server seems to be down or does not 
exist.
\end{verbatim}

Beyond these verification steps, using a logic programming approach
allows us to infer some important information. For example we found
out that 40 publications out of a total of 2816 were co-written by two
groups, giving an indicator on the way the 100 research groups
cooperate.

\begin{verbatim}
The citation  "Three knowledge representation 
formalisms for content-based manipulation of 
documents" line 2724 in file "acacia.xml" has 
been published in cooperation with  
orpailleur.
\end{verbatim}

Our system reported respectively 1372 and 1432 messages for the years
2001 and 2002. The reasons of this important number of errors are
various. There is a lot of misspelling in family names (mainly due to
missing accents or to differences between the name used in
publications and the civil name for women). Some persons who
participated to some parts of a project but have not been present
during all the year can be missing in the list of the team's
members. There is also a lot of mistakes in the list of publications:
a paper written during the current year can be published the year
after and should not appear as a publication of the year. There are
also a lot of errors in URLs and this number of errors should increase
as URLs are not permanent objects.

It is to be noticed that we have worked on documents that have been
generated, and that the original latex versions have been carefully
reviewed by several persons. This means that proofreading is not
feasible for large documents.

For future developments, other important indicators can also be
inferred: how many PhD students are working on the different teams,
how many PhD theses are published, etc. These indicators are already
used by the management to evaluate the global performance of the
institute but are compiled manually. An automatic process should raise
the level of confidence in these indicators.  They can also be
compared mechanically with other sources of information as, for example,
the data bases used by the staff office.

\section{Implementation notes}

All the implementation has been done in Prolog (more exactly Eclipse)
except the XML scanner which has been constructed with flex.

An XML parser has been generated using an extension of the standard
DCG. This extension, not yet published, gives some new facilities as
allowing left recursive rules and generating some efficient prolog
code. A particularity is that it constructs two resulting terms. The
first one is a parsed term, as usual. The second one is used to allow
a correspondence between an occurrence in the parsed term and the line
numbers in the source, allowing pertinent error messages as seen in
the previous section.

This parser has been extended to generate our specification language
parser.

The rule compiler as been entirely written in Prolog.

Concerning the execution of the specification, the main difficulty
comes from the fact that we have a global environment. The traditional
solution in this case is to use coroutines or delays. As our input
comes from many files, this solution was not reasonable, and we have
chosen a two pass process.  For each input file, during the first pass
we use the local environment and construct the part of the global
environment which is generated by the current file and a list of
delayed conditions which will be solved during the second pass. During
the second pass, all the individual parts of the global environment
are merged and the result is used to perform the delayed
verifications, producing errors messages when necessary.

\section{Conclusion}

In this paper, we have showed how techniques used to define the formal
semantics of programming languages can be used in the context of Web
sites. This work can be viewed as a complement to other researches
which may be very close: for example in~\cite{fernandez99verifying},
some logic programming methods are used to specify and check integrity
constraints in the structure of Web sites (but not to its static
semantics). Our work which is focused on the content of Web sites and
on their formal semantics, remains original. It can be extended to
the content management of more general information system.

As we have seen, the techniques already exist and are used in some
other domains. The use of logic programming, and in particular of
Prolog as an implementation language, is very well adapted to our
goals.  Our goal is to make these techniques accessible and easily
usable in the context of Web sites with the help of a specific
specification language.

We are now convinced that our technology is adequate.
We plan, in parallel with the development of our language, to explore
more deeply some applications, both on the verification side and the
inference side.

\section{Acknowledgment}

The author wants to thank Brigitte Trousse for its participation in an
earlier stage of this work.

\bibliographystyle{abbrv}

\appendix

\section{Our specification language\\syntax}

This appendix contains the source file of our parser. It uses a home
made extension of Define Clause Grammars. This extension has three
advantages: it handles left recursions; it takes advantage of prolog
hash-coding over some arguments; it produces a structure which is rich
enough to allow precise error messages with some back references to
the source that is analysed (it can be used in particular to retrieve
line numbers).

The production
\begin{verbatim}
attr(attr(N,V)) :- name(N), ['='], value(V).
\end{verbatim}
is equivalent to the traditional grammar rule
\begin{verbatim}
attr --> name '=' value.
\end{verbatim}
attr(N,V) is a regular prolog term and explains how the parse tree is
constructed from subtrees.

The call to {\tt stoken} is the interface with the lexer which is
implemented using flex.

Beside the specification language itself, we can recognize the syntax
of prolog terms (non terminal {\tt term}) and XML elements (non
terminal {\tt element}). Traditional elements are extended with
variables as in {\tt <\$varname>}. The sign {\tt <*} is used to
comment easily some complete rules.

\begin{verbatim}
entry(A) :- rules(A).

rules([]).
rules([A|B]) :- rule1(A), rules(B).

%-------

rule1(skip(A)) :- ['<*'], rule(A).
rule1(A) :- rule(A).

rule(ruleenv(A,B)) :- 
     left(A), ['=>'], right(B), [';'].
rule(ruletest(A,B)) :- 
     left(A), ['?'], test(B), [';'].

left(left(A,B)) :- 
     element(A), ['&'], cond_s(B).
left(left(A,[])) :- element(A).

cond_s([A|B]) :- cond(A), ['&'], cond_s(B).
cond_s([A]) :- cond(A).

cond(contains(A,B)) :- 
     variable(A), ['contains'], element(B).
cond(eq(A,B)) :- name(A), ['='], term(B).

right(right(A,B)) :- act(A), ['&'], right(B).
right(right(A,[])) :- act(A).

act(assign(A,B)) :- name(A), [':='], term(B).
act(A) :- term(A).

test(ifnot(A,B)) :- term(A), ['/'], conseq(B).
test(if(A,B)) :- term(A), ['->'], conseq(B).

conseq(A) :- element(A).
conseq(A) :- term(A).

%-------

term(term(Op,Args)) :- 
     name(Op), ['('], term_s(Args), [')'].
term(A) :- sstring(A).
term(var(A)) :- ['$'], name(A).

term_s([]).
term_s([A]) :- term(A).
term_s([A|Q]) :- term(A), [','], term_s(Q).

%-------
	
element(A) :- text(A).
element(empty_elem(A,T)) :-
     ['<'], name(A), attr_s(T), ['/>'].
element(elem(A,T,L)) :-
     ['<'], name(A), attr_s(T), ['>'],
     element_s(L),
     ['</'], name(B), ['>'].
element(var(A)) :- ['<$'], name(A), ['>'].

element_s([E|L]) :- element(E), element_s(L).
element_s([]).

attr_s([E|L]) :- attr(E), attr_s(L).
attr_s([]).

attr(attr(N,V)) :- name(N), ['='], value(V).

value(V) :- sstring(V).
value(V) :- variable(V).

variable(var(A)) :- ['$'], name(A).

% -----

sstring(string(V)) :- 
     stoken('STRING',string(V)).
sstring(string(V)) :- 
     stoken('STRING2',string(V)).
name(name(A)) :- stoken('NAME',string(A)).
text(text(A)) :- stoken('TEXT',string(A)).
\end{verbatim}

\section{Source for checking some part of an activity report}

This appendix contains the source file for our activity report
checker. This is the real source, and it may differ from what appears in
the text.

\begin{verbatim}
<raweb>
  <accueil>
    <$_> 
    <$_>
    <projet><$P><$_></projet>
    <$_>
  </accueil>
  <$_>
</raweb>
	=> project := $P;

<raweb year=$X> <$_> </raweb> 
	=> year := $X & defperso := "false";

<catperso> <$_> </catperso>
	=> defperso := "true";

<pers prenom=$P nom=$N> <$_> </pers>
& defperso = "true"
& project = $Proj
	=> personne($P,$N,$Proj);

<pers prenom=$P nom=$N> <$_> </pers>
& defperso= "false"
& project = $Proj 
? personne1($P,$N,$Proj) / 
	<li>
	  Warning: <i> <$P> <$N> </i>
	  does not appear in the list of project's 
          members; line <$SourceLine> in 
          <$SourceFile>.
	<p> </p>
	</li> ;


<citation from=$X> <$_> </citation>
	=> citationfrom := $X;

<citation> <$A> </citation>
& $A contains <btitle> 
                <$Title> 
                <$_> 
              </btitle>
=> title := $Title ;

<byear> <$Byear> <$_>  </byear>
& citationfrom = "year"
& year = $Year
& title = $Title
? sameyear($Byear,$Year) /
	<li>
	  Warning: The citation <i> "<$Title>" </i> 
          line <$SourceLine>
	  in file
	  <$SourceFile>
	  has not been published during this year
	  (published in <$Byear>).
	<p> </p>
	</li> ;

<btitle> <$Title> <$_> </btitle>
& citationfrom = "year"
& project = $Proj 
	=> pub($Title,$Proj) ;


<btitle> <$Title> <$_> </btitle>
& citationfrom = "year"
& project = $Proj 
? pubbyotherproject($Title,$Proj,$Otherproj) 
  ->
	<li>
	 Hourra! the citation <i> "<$Title>" </i> 
          line
	  <$SourceLine>
	  in file
	  <$SourceFile>
	has been published in cooperation with 
	<$Otherproj>.
	<p> </p>
	</li> ;

<xref url=$URL><$_></ref>
? testurl($URL,$Answer1,$Answer2) ->
	<li>
	Testing of URL <i> <$URL> </i> line
	  <$SourceLine>
	  in file
	  <$SourceFile> replies:
	<$Answer1>
	<$Answer2>.
	<p> </p>
	</li> ;
\end{verbatim}

The predicates {\tt personne1}, {\tt sameyear}, {\tt
pubbyotherproject}, and {\tt testurl} are defined directly in
Prolog. The last one makes a call to the program {\tt wget} with some
timeout guard.

\balancecolumns 

\end{document}